\documentclass[]{article}
\usepackage[ruled,vlined]{algorithm2e} 
\usepackage{amssymb}
\usepackage{graphicx}
\usepackage{tikz}
\usetikzlibrary{arrows,calc,positioning}
\usetikzlibrary{automata}
\usepackage{pgfplots}
\pgfplotsset{compat=newest}
\usetikzlibrary{plotmarks}
\usepackage{grffile}
\usepackage{amsmath,amssymb}
\usetikzlibrary{shapes.gates.logic.US,shapes.gates.logic.IEC,calc}
\usepackage{url}
\usepackage{pdflscape}
\usepackage{afterpage}
\usepackage{psfrag}
\usepackage{epstopdf}
\usepackage{caption}
\DeclareCaptionType{copyrightbox}
\usepackage{ifthen}
\usepackage{subcaption} 
\usepackage{cleveref}
\usetikzlibrary{pgfplots.groupplots}
\usepackage{varwidth}
\newenvironment{pseudocode}[1][htb]
{%
	\begin{algorithm}[#1]%
	}{\end{algorithm}}
\urldef{\mailsa}\path|{nathalie.cauchi, khaza.hoque, aabate}@cs.ox.ac.uk|  

\newcommand{\po}{\mathrm{P}}     
\newcommand{\ro}{\mathrm{R}}     
\newcommand{\fo}{\mathrm{F}}     
\newcommand{\Rl}{\mathbb{R}}     
\newcommand{\Rt}{\textbf{R}}	 
\newcommand{\Na}{\mathbb{N}}     
\newcommand{\F}{$\mathcal{F}$}   
\newcommand{\Tcln}{T_{cln}}      
\newcommand{\Trpl}{T_{rplc}}     
\newcommand{\Td}{T_{deg}}        
\newcommand{\Toh}{T_{oh}}        
\newcommand{\Tin}{T_{insp}}        
\newcommand{\Trp}{T_{rep}}       
\newcommand{\Act}{\text{TL}}    
\newcommand{\Ap}{\text{AP}}		 
\newtheorem{remark}{Remark}
\newtheorem{definition}{Definition}
\newtheorem{example}{Example}

\definecolor{mycolor1}{rgb}{0.00000,0.44700,0.74100}%
\definecolor{mycolor2}{rgb}{0.85000,0.32500,0.09800}%
\definecolor{mycolor3}{rgb}{0.92900,0.69400,0.12500}%
\definecolor{mycolor4}{rgb}{0.49400,0.18400,0.55600}%
\definecolor{mycolor5}{rgb}{0.46600,0.67400,0.18800}%
\definecolor{mycolor6}{rgb}{0.30100,0.74500,0.93300}%
\definecolor{mycolor7}{rgb}{0.63500,0.07800,0.18400}
\usepackage{authblk}
\usepackage{blindtext}
\title{\textbf{Efficient Probabilistic Model Checking of Smart Building Maintenance using Fault Maintenance Trees}}
\author[1]{\normalsize Nathalie Cauchi}
\author[1]{ Khaza Anuarul Hoque}
\author[1]{ Alessandro Abate}
\author[2]{ Mari\"{e}lle Stoelinga}
\affil[1]{ Department of Computer Science, University of Oxford, Oxford UK}
\affil[1]{\textit {\{name.surname\}@cs.ox.ac.uk}}
\affil[2]{ Formal Methods and Tools Group, University of Twente, Twente, The Netherlands}
\affil[2]{\textit {marielle@cs.utwente.nl}}
\date{}

\begin{document}

\maketitle

\begin{abstract}
		{Cyber-physical systems, like Smart Buildings and power plants, 
	have to meet high standards, both in terms of reliability and availability. Such metrics are typically evaluated using Fault trees (FTs) and do not consider maintenance strategies which can significantly improve lifespan and reliability. 
}
Fault Maintenance trees (FMTs) -- an extension of FTs that also incorporate maintenance and degradation models, are a novel technique that serve as a good planning platform for balancing total costs and dependability of a system. In this work, we apply the FMT formalism to a Smart Building application. 
We propose a framework for modelling  FMTs using probabilistic model checking and present an algorithm for performing abstraction of the FMT in order to reduce the size of its equivalent Continuous Time Markov Chain. This allows us to apply the probabilistic model checking more efficiently. We demonstrate the applicability of our proposed approach by evaluating various dependability metrics and maintenance strategies of a Heating, Ventilation and Air-Conditioning system's FMT.
\end{abstract}

	\section{Introduction}
\label{sec:Intro}

Worldwide, buildings account for approximately 40\% of the total energy consumption and 20\% of the total $CO_2$ emissions, annually~\cite{Directive2010}. 
Efficient Building Automation Systems (BAS) can reduce energy consumption by up to 30\%  through their optimal operation, continuous commissioning and maintenance~\cite{Directive2010}. Constructions employing such technologies are termed \textit{Smart Buildings}. High standards have to be adhered by such technologies,  both  in terms  of  reliability and  availability. 	
One way of achieving this is by employing methods
to perform preventative and predictive maintenance actions. 
Diagnostic and fault detection techniques for Smart Building applications have been developed in 
~\cite{Yan2017,boem2017distributed}. Predictive and preventative maintenance strategies are devised in~\cite{cauchi2017model,Babishin201610480}. 
However, these techniques preclude availability and reliability measurements and focus only on synthesis of maintenance policies in the presence of degradation and fault finding.
Reliability and availability are typically tackled using Fault Trees (FTs), where the focus is on finding the root causes of a system failure using a top-down approach.
FTs do not include maintenance strategies in the analysis -- a key element in reducing component failures. \cite{ruijters2016fault} presents the Fault Maintenance Tree (FMT) as 
an extension of FT encompassing both degradation and maintenance models. The degradation models represent the different levels of component degradation and are known as Extended Basic Events (EBE). The maintenance models incorporate the undertaken maintenance policy which includes both inspections and repairs. These are modelled using Repair and Inspection modules in the FMT framework. 

In literature, FMTs are analysed using Statistical Model Checking technique (SMC)~\cite{ruijters2016fault}
and provide statistical guarantees. In contrast, Probabilistic Model Checking (PMC), based on numerical analysis, provide formal guarantees with higher accuracy when compared with SMC~\cite{younes2006numerical}. However, numerical methods are far more memory intensive and may result in a state space  explosion. This limitation of PMC often leaves SMC as the last resort~\cite{younes2006numerical}.
In this paper we tackle the FMT analysis using PMC. 
Our contributions can be summarised as follows: 
\begin{enumerate}
	\item We formalise the FMT framework using Continuous Time Markov Chain (CTMCs). 
	\item We formalise the dependability metrics 
	using the extended Continuous Stochastic Logic (CSL) formalism such that they can be computed using the PRISM model checker~\cite{PRISM}.
	\item To mitigate the state space explosion problem, we present an FMT abstraction technique which decomposes a large FMT into an equivalent abstract  FMT based on our proposed graph decomposition algorithm. 
	Using our framework, we are able to achieve a 67$\%$ reduction in the state space  size.
	\item Finally, we construct a FMT that identifies failure of a Heating, Ventilation and Air-conditioning system (HVAC). We apply the developed framework to the built FMT and evaluate relevant dependability metrics, together with different maintenance strategies using the PRISM model checker. 
\end{enumerate}

To the best of our knowledge, this is the first attempt to analyse FMTs using Probabilistic Model Checking and also the first application to Smart Building systems.

{This article has the following structure: 
	Section~\ref{sec:Prelim} introduces the fault maintenance trees and probabilistic model checking frameworks. This is 
	followed by the developed methodology for modelling FMT using CTMCs and performing model checking in Section~\ref{subsec:FMTFrame:Modelling}. 
	The framework is applied to a heating, ventilation and air-conditioning (HVAC) case study which is presented in Section~\ref{sec:CaseStudy}.}
	\section{Preliminaries}
\label{sec:Prelim}
\subsection{Fault maintenance trees framework}
\label{subsec:Prelim:FMTFrame}

Fault trees are directed acyclic graphs (DAG) describing the combinations of component failures that lead to system failures. The leaves in the fault trees are called \textit{basic events} and denote the system failures. The internal nodes of the graph are called \textit{gates} and describe the different ways that failures can interact to cause other components to fail. The gates in a fault tree can be of several types and these include the AND gate, OR gate, k/N-gate~\cite{ruijters2016fault}. 

Fault maintenance trees (FMT) extend fault trees by including maintenance (all the standard FT gates are also employed by the FMTs). This is achieved by making use of: 
\begin{enumerate}
	\item \textit{Extended Basic Events} - The basic events are modified to incorporate degradation models of the component the leaf represents. The degradation models represent different discrete levels of degradations the components can be in and are a function of time.
	\item \textit{Rate Dependency Events} - A new gate introduced in ~\cite{ruijters2016fault}, labelled as \textit{RDEP} that accelerates the degradation rates of dependent child nodes and is depicted in Figure~\ref{fig:RDEPGate}.
	When the component connected to the input of the RDEP fails, the degradation rate of the dependent components is accelerated  with an acceleration factor $\gamma$.  
	\begin{figure}[ht!]
		\centering
		\resizebox{3cm}{!}{ \begin{tikzpicture}[
      >=latex',
      auto
    ]
  \tikzstyle{int}=[draw,minimum size=2.5em,text centered,text width=3.5cm]
  \tikzstyle{intg}=[draw,minimum size=2.5em,text centered,text width=3.5cm]
     \node [int]  (ki1) [node distance=1.5cm] {\LARGE \textbf{RDEP}};
      \node [intg] (ki3) [node distance=.8cm,below of=ki1] { };
      \draw (-1.875,0) node[anchor=north]{}
  -- (-1.875,-.25) node[anchor=north]{}
  -- (-1.5,-0.125) node[anchor=south]{}
  -- cycle;
     \draw (-1.875,-0.15) node[anchor=north]{}
  -- (-3.375,-0.15) node[anchor=north]{\large input}
  -- cycle;
       \draw (0,-1.1) node[anchor=north]{}
  -- (0,-2) node[anchor=north]{\large Children (n)}
  -- cycle;
  
    \end{tikzpicture}}
		\caption{{RDEP gate with 1 input and dependent components also known as children.}}
		\label{fig:RDEPGate}
	\end{figure}
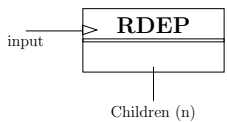
	\item \textit{Repair and Inspection modules} - 
	The repair module (RM) performs cleaning or replacements actions. These actions can be either carried out using fixed time schedules or when enabled by the inspection module (IM). The IM performs periodic inspections and when components fall below a certain degradation threshold a repair or partial replacement is initiated by the IM to be performed by the RM.
\end{enumerate}
\subsection{Probabilistic model checking}
\label{subsec:Prelim:PMC}
Model checking 
is a well-established formal verification technique used to verify the correctness of finite-state systems. Given a formal model of the system to be verified in terms of labelled state transitions and the properties to be verified in terms of temporal logic, the model checking algorithm exhaustively and automatically explores all the possible states in a system to verify if the property is satisfiable or not. 
\emph{Probabilistic model checking} deals with systems that exhibit stochastic behaviour
and is based on the construction and analysis of a probabilistic model of the system. We make use of CTMCs{, having both transition and state labels, } to perform stochastic modelling. Properties are expressed in the form of 
extended Continuous Stochastic Logic (CSL)~\cite{kwiatkowska2007stochastic}. 

\begin{definition}[Continuous time Markov chain (CTMC)] \label{def:CTMC}%
	{The tuple $C =(S, s_0, \Act, \Ap, L, \Rt)$ defines a CTMC which is composed of a set of states $S$, the initial state $s_0$, a finite set of transition labels $\Act$, a finite set of atomic propositions $\Ap$, a labelling function $L: S\rightarrow 2^{AP}$ and  the transition rate matrix $ \Rt: S \times  S \rightarrow \Rt_{\geq 0}$. }The rate $\Rt(s,s')$ defines the delay before which a transition between states $s$ and $s'$ takes place. If $\Rt(s,s') \ne 0$ then the probability that a transition between the states $s$ and $s'$ is 
	defined as $1 - e^{-{{\Rt}}(s,s') t} $ where $t$ is time. No transitions will  trigger if $\Rt(s,s') = 0$. 
	
\end{definition}
The logic of  CSL specifies state-based properties for CTMCs, built out of propositional logic
, a steady-state operator that refers to the stationary probabilities, and a probabilistic operator for reasoning about transient state probabilities. The state formulas are interpreted over states of a CTMC, whereas the path formulas are interpreted over paths in a CTMC. For detail about the syntax and semantics of CSL (which also includes reward formulae), we refer the interested readers to~\cite{kwiatkowska2007stochastic}.

Examples of a CSL property with its natural language translation are: (i) $ \po_{\geq 0.95} [\fo ~complete]$ - ``The probability of the system eventually completing its execution successfully is at least 0.95". 
(ii) $\ro_{=?} [\fo ~success] $ - ``What is the expected reward accumulated before the system successfully terminates?" 


\section{Formalizing FMTs using CTMCs}
\label{subsec:FMTFrame:Modelling}
In this section, we 
first formalise the FMT framework by presenting the formal syntax and semantics for modelling FMTs using CTMCs. 
Next, we list the set of metrics used to analyse the FMT. Finally, we present the developed framework which allows us to analyse large FMTs using probabilistic model checking (PMC).  
	\subsection{FMT Syntax}
\label{subsubsec:FMTFrame:Modelling:Syntax}
To formalise the syntax of FMTs using CTMCs, we first define the set \F, characterizing each FMT element by {type, inputs and  rates}. We introduce a new element called DELAY which will be used to model the deterministic time delays required by the extended basic events (EBE), repair module (RM) and inspection module (IM).  We restrict the set \F to contain the  EBE, RDEP gate, OR gate, DELAY, RM and IM modules since these will be the components used in the case study presented in Section \ref{sec:CaseStudy}. 

\begin{definition}[Elements of fault maintenance tree]\label{def:FMTSem}
	The set \F of FMT elements consists of the following tuples. Here, $n, N \in \Na$ are natural numbers, $\mathit{thresh}, in, \mathit{trig} \in \{0,1\}$ take binary values, {$\Td,\Tcln$, $\Trpl,$ $\Trp,\Toh$, $\Tin \in \Rl^{\ge 0}$ are deterministic delays, and $\gamma \in \Rl^{\ge 0}$ is a rate}.
	\begin{itemize}		
		\item $(EBE,\Td,\Tcln,\Trpl,N)$ represent the extended basic events  with $N$ discrete degradation levels, each of which degrade with a time delay equal to $T_{deg}$. It also takes as inputs the time taken to restore the EBE to the previous degradation level $\Tcln$ when cleaning is performed and the time taken to restore the EBE to its initial state $\Trpl$ following a replacement action.
		\item $(RDEP,n,\gamma,in, \Td)$ represents the RDEP gate with $n$ dependent children, acceleration rate $\gamma$, the input $in$ which activates the gate and $\Td$ the degradation rate of the dependent children.
		\item $(OR,n)$ represents the OR gate with $n$ inputs.
		\item $(RM,n,\Trp,\Toh,\Tin,\Tcln, \Trpl,{\mathit{thresh}}, \mathit{trig})$ represents the RM module which acts on $n$ EBEs (in our case, this corresponds to all the EBEs in the FMT).	The RM can either be triggered periodically to perform a cleaning action, every $\Trp$ delay, or a replacement action, every $\Toh$ delay, or by the IM when the delay $\Tin$ has elapsed and the $\mathit{thresh}$ condition is met. The time to perform a cleaning action is $\Tcln$, while the time taken to perform a replacement is $\Trpl$. The $\mathit{trig}$ signal ensures that when the component is not in the degraded states, no unnecessary maintenance actions are carried out. 
		\item $(IM,n,\Tin,\Tcln,\Trpl,\mathit{thresh})$ represents the IM module which acts on $n$ EBEs (in our case, this corresponds to all the EBEs in the FMT). The IM initiates a repair depending on the current state of the EBE. Inspections are performed in a periodic manner, every $\Tin$. If during an inspection, the current state of the EBE does not correspond to the \textit{new} or \textit{failed} state (i.e. the degradation level of the inspected EBE is below a certain threshold), the $thresh$ signal is activated and is sent to the RM. Once a repair action is performed the IM moves back to the initial state with a delay equal to  $\Tcln$ or $\Trpl$ depending on the maintenance action performed. 
		\item $(DELAY, T, N)$ represents the DELAY module which takes two inputs representing the deterministic delay $T \in \{\Td,$ $\Tcln,\Trpl,\Trp,\Toh,\Tin\}$ to be approximated using an Erlang distribution with $N$ number of states. 
		This DELAY module can be extended 
		by inclusion of a reset transition label, which when triggered restarts the approximation of the deterministic delay before it has elapsed. The extended DELAY module is referred to as  $(DELAY, T, N)_{ext}$.
	\end{itemize}
\end{definition}
The FMT is defined as a special type of directed acyclic graph $G=(V,E)$ where the vertices $V$ represent the gates and the events which represent an occurrence within the system, typically the failure of a subsystem down to an individual component level, and the edges $E$ which represent the connections between vertices. Events can either represent the EBEs or \textit{intermediate} events which are caused by one or more other events. The event at the top of the FMT is the top event (TE) and corresponds to the event being analysed -  modelling the failure of the (sub)system under consideration. The EBE are the leaves of the DAG. 
For $G$ to be a well-formed FMT, we take the following assumptions 
(i) vertices are composed of the OR, RDEP gates, (ii) there is only one top event, (iii) RDEP can only be triggered by EBEs and (iv) RM and IM are not part of the DAG tree but are modelled separately \footnote{Note, for different FMT structure same RM and IM modules are used, thus RM and IM modules are independent of FMT structure}. This DAG formulation allows us to propose a framework in Subsection~\ref{subsec:FMTFrame:Dec} such that we can efficiently perform probabilistic model checking. 

\begin{definition}[Fault maintenance tree] \label{def:FMT}
	A fault maintenance tree is a directed acyclic graph 
	\noindent $G = (V,E)$ composed of vertices $V$ and edges $E$.	
\end{definition}

\subsection{Semantics of FMT elements }
\label{subsubsec:FMTFrame:Modelling:SemanticsElements}

Next, we provide the CTMC semantics for each FMT element $f \in \mathcal{F}$. These elements are then instantiated based on the underlying FMT structure to form the semantics of the whole FMT in CTMC form.

\begin{figure}[ht!]
	\centering
	\resizebox{\textwidth}{!}{\begin{tikzpicture}[font=\sffamily]

\tikzset{node style/.style={state, 
		minimum width=1.2cm,
		line width=0.5mm,
		fill=white!20!white}}

\node[node style, initial above] at (0, 0)  (s0)     {\Large $s_0$};
\node[node style] at (5, 0)  (s1)     {\Large $s_1$};
\node[node style] at (10, 0) (s2) {\Large $s_2$};
\node[node style] at (15, 0) (s3) {\Large $s_3$};
\draw[every loop,
auto=right,
line width=0.5mm,
>=latex,
draw=black,
fill=black]
(s0)     edge[bend left=20, auto=left] node {$\texttt{degrade}_1,\lambda$} (s1)
(s1)     edge[bend left=20, auto=left] node {$\texttt{degrade}_2,\lambda$} (s2)
(s2)     edge[bend left=20, auto=left] node {$\texttt{degrade}_3,\lambda$} (s3)
(s3)     edge[bend left=10, auto=right] node {$\texttt{perform\_clean},1$} (s2)
(s2)     edge[bend left=10, auto=right] node {$\texttt{perform\_clean},1$} (s1)
(s1)     edge[auto=right] node {$\texttt{perform\_clean},1$} (s0)
(s3)     edge[bend left=23, auto=right] node {$\texttt{perform\_replace},1$} (s0)
(s2)     edge[bend left=23, auto=right] node {$\texttt{perform\_replace},1$} (s0)
(s1)     edge[bend left=23, auto=right] node {$\texttt{perform\_replace},1$} (s0);
\end{tikzpicture}}
	\caption{{CTMC representing the EBE with $N =3$ with the transition labels $\Act_{EBE} =\{ \texttt{degrade}_{i \in \{1,2,3\}}, $ $\texttt{perform\_clean}, \texttt{perform\_replace}\}$  on each of the state transitions. The state labels are not shown and the initial state is pointed to by the arrow labelled with start.}} 
	\label{fig:EBE}
\end{figure}
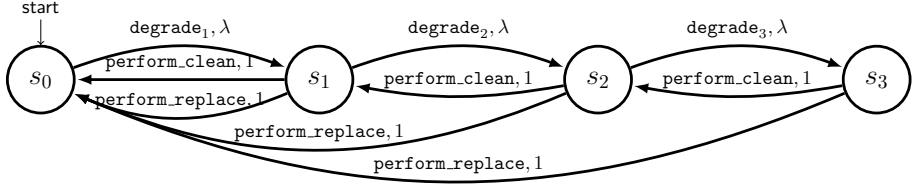

\paragraph{\textbf{DELAY}}
\label{par:FMTFrame:Modelling:Clck}
{We define the semantics for the $(DELAY,T,N)$ element 
	using Figure~\ref{fig:ClkNoRst} and describe the  corresponding CTMC  
	using the set of states given by  $D =\{d_0,d_1,\dots, d_{N+1}\}$, the initial state $d_0$, the set of transitions labels  $\Act=\{\texttt{trigger}, \texttt{move}\}$, the set of atomic propositions $\Ap =\{T\}$ with 
	$L(d_0)  = \dots =L(d_{N}) = \emptyset$, and $L(d_{N+1}) = \{T\}$. The rate matrix $\Rt$ becomes clear from Figure~\ref{fig:ClkNoRst} and  
	\begin{align}
	\Rt_{ij} = \begin{cases}
	\mu &  i= 0 \wedge j =1\\
	\frac{N}{T} & ((i \ge 1 \vee i < N+1)  \wedge  j = i+1) \\
	&\vee (i=N+1 \wedge j=1) \\
	0 & \text{otherwise},
	\end{cases}
	\end{align}
	with $i$ representing the current state, $j$ is the next state and $\mu$ is a fixed large value corresponding to introducing a negligible delay, which is used to trigger all the DELAY modules at the same time (cf. Definition~\ref{def:CTMC}).  }
In Figure~\ref{fig:ClkRst} we define the semantics of $(DELAY,T,N)_{ext}$. This results in the CTMC described using the state space $D =\{d_0,d_1,\dots, d_{N+1}\}$, the initial state $d_0$, the set of transition labels $\Act=\{\texttt{trigger},$ $ \texttt{move}, \texttt{reset}\}$, the set of 
atomic propositions $\Ap =\{T\}$, the labelling function $L(d_0) = L(d_1) = \dots =L(d_{N}) = \emptyset$, and $L(d_{N+1}) = \{T\}$ and the rate matrix $\Rt$ where \begin{align}
\Rt_{ij} = \begin{cases}
\mu &  i= 0 \wedge j =1\\
1           & (i \ge 2 \vee i < N+1) \wedge  j = 1\\
\frac{N}{T} & ((i \ge 1 \vee i < N+1)  \wedge j = i+1)\\
& \vee (i=N+1 \wedge j=1) \\
0 & \text{otherwise},
\end{cases}
\end{align}
with $i$ representing the current state and $j$ is the next state. In both instances, the deterministic delays is approximated using an Erlang distribution
~\cite{hoque2015towards} and all DELAY modules are synchronised to start together using the \texttt{trigger} transition label.  The extended DELAY module have the transition labels \texttt{reset} which restarts the Erlang distribution approximation whenever the guard condition is met at a rate of $1 \times \Rt_{sync}$ where $\Rt_{sync}$ is the rate coming from the use of  synchronisation with other modules causing the reset to occur ( as explained in Subsection~\ref{subsubsec:FMTFrame:Modelling:Semantics}). This is required when a maintenance action is performed which restores the EBE's state back to the original state and thus restart the degradation process, before the degradation time has elapsed. 
\begin{remark} \label{remark:Er}
	The basic properties of an Erlang distribution: A random variable $Z \in \Rl_{+}$ has an Erlang distribution with 	$k \in \Na$ stages and a rate $\lambda \in \Rl_{+}, Z \sim Erlang(k,\lambda)$, if $Z = Y_1 + Y_2 + \dots Y_k$ where each $Y_i$ is exponentially distributed with rate $\lambda$. The cumulative density function of the Erlang distribution is characterised using,
	\begin{equation} \label{eqn:cdf}
	f(t;k,\lambda) = 1- \sum_{n=0}^{k-1}\frac{1}{n!} \exp(-\lambda t)(\lambda t)^n \quad \textrm{for } t,\lambda \ge 0
	\end{equation}
	and for $k=1$, the Erlang distribution simplifies to the exponential distribution. In particular, the sequence $Z_k \sim Erlang(k,\lambda k)$ converges to the deterministic value $\frac{1}{\lambda}$ for large $k$. Thus, we can approximate a deterministic delay $T$ with a random variable $Z_k \sim Erlang(k,\frac{k}{T})$~\cite{bortolussi2012fluid}. Note, there is a trade-off between the accuracy and the resulting blow-up in size of the CTMC model for larger values of $k$ (a factor of $~k$ increase in the model size)~\cite{hoque2015towards,hoque2014probabilistic}. In this work, the Erlang distribution will be used to model the fixed degradation rates, the maintenance and inspection signals. This is a similar approach taken in ~\cite{ruijters2016fault} where degradation phases are approximated by an (k,$\lambda$)-Erlang distribution.
\end{remark}
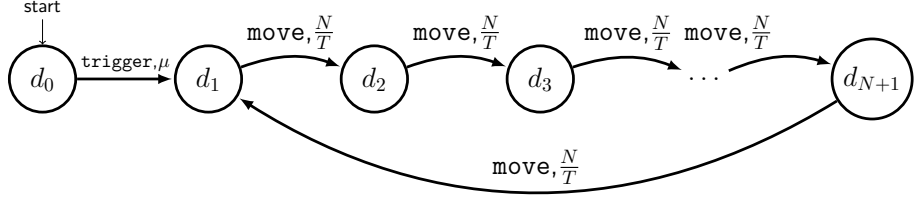
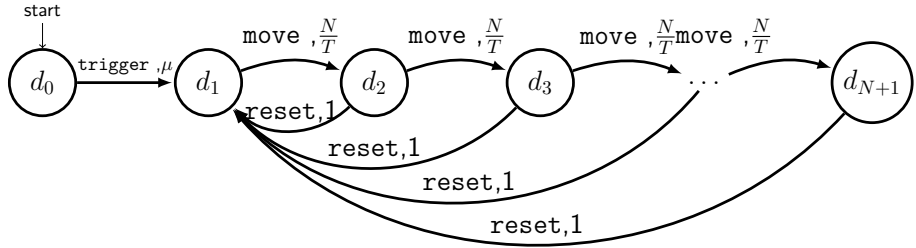
\begin{figure}[ht!]
	\psfragscanon
	\begin{subfigure}{\textwidth}
		\resizebox{\textwidth}{!}{\begin{tikzpicture}[font=\sffamily]

\tikzset{node style/.style={state, 
		minimum width=1.2cm,
		line width=0.5mm,
		fill=white!20!white}}

\node[node style, initial above] at (0, 0)  (s0)     {\Large $d_0$};
\node[node style] at (3, 0)  (s1) {\Large $d_1$};
\node[node style] at (6, 0)  (s2) {\Large $d_2$};
\node[node style] at (9, 0)  (s3) {\Large $d_3$};
\node[]           at (12, 0) (s4) {\Large $\dots$};
\node[node style] at (15, 0) (s5) {\Large $d_{N+1}$};
\draw[every loop,
auto=right,
line width=0.5mm,
>=latex,
draw=black,
fill=black]
(s0)     edge[ auto=left] node {\texttt{trigger},$\mu$} (s1)
(s1)     edge[bend left=20, auto=left] node {\Large\texttt{move},$\frac{N}{T}$} (s2)
(s2)     edge[bend left=20, auto=left] node {\Large\texttt{move},$\frac{N}{T}$} (s3)
(s3)     edge[bend left=20, auto=left] node {\Large\texttt{move},$\frac{N}{T}$} (s4)
(s4)     edge[bend left=20, auto=left] node {\Large\texttt{move},$ \frac{N}{T}$} (s5)
(s5)     edge[bend left=32, auto=right] node {\Large{\texttt{move}},\Large$\frac{N}{T}$} (s1);
\end{tikzpicture}}
		\caption{{CTMC representing DELAY with $N$ states used to approximate a delay equal to $T$ approximated using $Erlang(N,\frac{N}{T})$. The transition labels $\Act =\{ \texttt{trigger}, $ $\texttt{move}\}$ are shown on each of the transitions. The state labels are not shown and the initial state of the CTMC is pointed to using an arrow labelled with start. }}\label{fig:ClkNoRst}
	\end{subfigure}
	\\
	\begin{subfigure}{\textwidth}
		\resizebox{\textwidth}{!}{\begin{tikzpicture}[font=\sffamily]

\tikzset{node style/.style={state, 
		minimum width=1.2cm,
		line width=0.5mm,
		fill=white!20!white}}

\node[node style, initial above] at (0, 0)  (s0)     {\Large $d_0$};
\node[node style] at (3, 0)  (s1)     {\Large $d_1$};
\node[node style] at (6, 0)  (s2) {\Large $d_2$};
\node[node style] at (9, 0)  (s3) {\Large $d_3$};
\node[]           at (12, 0) (s4) {\Large $\dots$};
\node[node style] at (15, 0) (s5) {\Large $d_{N+1}$};

\draw[every loop,
auto=right,
line width=0.5mm,
>=latex,
draw=black,
fill=black]
(s0)     edge[ auto=left] node {\texttt{trigger} ,$\mu$} (s1)
(s1)     edge[bend left=20, auto=left] node {\Large\texttt{move} ,$\frac{N}{T}$} (s2)
(s2)     edge[bend left=20, auto=left] node {\Large\texttt{move} ,$\frac{N}{T}$} (s3)
(s3)     edge[bend left=20, auto=left] node {\Large\texttt{move} ,$\frac{N}{T}$} (s4)
(s4)     edge[bend left=20, auto=left] node {\Large\texttt{move} ,$\frac{N}{T}$} (s5)
(s2)     edge[bend left=44, auto=right] node {\Large\texttt{reset},1} (s1)
(s3)     edge[bend left=46, auto=right] node {\Large\texttt{reset},1} (s1)
(s4)     edge[bend left=48, auto=right] node {\Large\texttt{reset},1} (s1)
(s5)     edge[bend left=48, auto=right] node {\Large\texttt{reset},1} (s1);
\end{tikzpicture}}
		\caption{{CTMC representing the extended DELAY with $N$ states used to approximate a delay equal to $T$. Delay approximated using $Erlang(N,\frac{N}{T})$. The transition labels $\Act =\{ \texttt{trigger}, $ $\texttt{move}, \texttt{reset}\}$ are shown on each of the state transitions, while the state labels are not shown. 
		}}\label{fig:ClkRst}
	\end{subfigure}
	\caption{{CTMC for (a) DELAY and (b) DELAY with reset guard.}}
	\label{fig:Clk}
\end{figure}
\paragraph{\textbf{Extended Basic Events (EBE)}}
\label{par:FMTFrame:Modelling:EBE}
The EBE 
are the leaves of the FMT and incorporate the component's degradation model. EBE are a function of the total number of degradation steps $N$ considered. 
Figure \ref{fig:EBE} shows the semantics of the $(EBE,T_{deg}, T_{cln},T_{rep},N=3)$. The corresponding CTMC {is described by the tuple $(\{s_0,s_1,s_2,s_3\},$ $ s_0,\Act_{EBE},$  $\Ap_{EBE}, L_{EBE}, \Rt_{EBE})$ where  $s_0$ is the initial state , \begin{align*}
	\Act_{EBE} &=\{ \texttt{degrade}_{i\in\{0,\dots,N\}}, \texttt{perform\_clean},\texttt{perform\_replace}\},
	\end{align*} the atomic propositions $\Ap_{EBE} = \{\mathit{new},\mathit{thresh},$ $ \mathit{failed}\}$, the labelling function $L(s_0) = \{new\},$ $ L(s_1) = L(s_2) = \{thresh\}, L(s_3) = \{failed\}$ }and 	$
\Rt_{EBE}= \left[\begin{smallmatrix}
0&1&0&0\\
1&0&1&0\\	
1&1&0&1\\
1&0&1&0\\
\end{smallmatrix}\right].$
{The deterministic time delays taken as inputs are modelled using three different DELAY modules:
	\begin{enumerate}
		\item an extended DELAY module approximating $\Td$ with the transition label \texttt{move} replaced with \texttt{degrade}$_N $such that synchronisation between the two CTMCs is performed (explained in Subsection~\ref{subsubsec:FMTFrame:Modelling:Semantics}).  When $\Td$ has elapsed the transition labelled with 
		\texttt{degrade}$_N$ is triggered and the EBE moves to the next state at a rate  equal to $\frac{N}{\Td} \times 1$\footnote{This is a direct consequence of synchronisation and corresponds to $\Rt \times \Rt_{EBE}$. Refer to Subsection~\ref{subsubsec:FMTFrame:Modelling:Semantics}}. The \texttt{reset} transition label and corresponding transitions are replicated in extended DELAY module and replaced with \texttt{perform\_clean} and \texttt{perform\_replace}. When the corresponding maintenance action is performed one of the transition label is triggered and the state of the EBE moves to previous state (if cleaning action is carried out) or to the initial state (if replace action is performed).
		\item a DELAY module approximating $\Tcln$  with the transition label \texttt{move} replaced with \texttt{perform\_clean}.  When $\Tcln$ has elapsed the transition with transition label \texttt{perform\_clean} is triggered and the EBE moves to the previous state at a rate  equal to $\frac{N}{\Tcln}$. 
		\item a DELAY module approximating $\Trpl$  with the transition label \texttt{move} replaced with \texttt{perform\_replace}.  When $\Trpl$ has elapsed the transition having the transition label \texttt{perform\_replace} is triggered and the EBE moves to the initial state at a rate  equal to $\frac{N}{\Trpl}$. 
	\end{enumerate}
	The transition labels \texttt{perform\_clean} and \texttt{perform\_replace} cannot be triggered at the same time and it is assumed that $\Tcln \neq \Trpl$. This is a realistic assumption as only one maintenance action is performed at the same time. } 	

\paragraph{\textbf{RDEP gate}}
\label{par:FMTFrame:Modelling:RDEP}
The RDEP gate has static semantics and is used in combination with the semantics of its $n$ dependent EBEs. When triggered ($in=1$), the associated EBE reaches the state labelled $\mathit{failed}$, the degradation rate of the $n$ dependent children is accelerated by a factor $\gamma$. We model the $in$ signal using,
\begin{align}\label{eqn:in}
in &= \begin{cases}
1 & L(s) = \mathit{failed},\\
0 & \text{otherwise},\\
\end{cases}
\end{align}
where $L(s)$ is the label of the  current state of the associated EBE. 
{ Similarly, we map the RDEP gate function using,
	\begin{align}\label{eqn:rdep}
	RA &= \begin{cases}
	\gamma T_{deg_1},\dots, \gamma T_{deg_n}& in = 1,\\
	T_{deg_1},\dots, T_{deg_n} & \text{otherwise},\\
	\end{cases}
	\end{align}
	where $T_{deg_i}, i \in {1, \dots n}$ corresponds to the degradation rate of the $n$ dependent children.}   \footnote{Note, this effectively results in changing the deterministic delay being modelled by the DELAY module to a new value if $in = 1$.} 

\paragraph{\textbf{OR gate}}
\label{par:FMTFrame:Modelling:OR}
The OR gate 
indicates a failure when either of its input nodes have failed and also does not have semantics itself but is used in combination with the semantics of its $n$ dependent input events  (EBEs or intermediate events). 
We use,
{\begin{align}\label{eqn:OR}
	FAIL &= \begin{cases}
	0    & E_{1} = 1 \wedge \dots \wedge E_{n} = 1 \\
	1    & \text{otherwise}\\
	\end{cases}
	\end{align}}
where $E_{i} = 1, i \in 1 \dots n$ corresponds to when the $n$ events, connected to the OR gate, represent a failure in the system. In the case of EBEs, $E_1 =1$ occurs when the EBE reaches the $\mathit{failed}$ state .

\begin{figure}[ht!]
	\psfragscanon
	\begin{subfigure}{\textwidth}
		\centering
		\resizebox{0.7\textwidth}{!}{    \begin{tikzpicture}[font=\sffamily]
 
        \tikzset{node style/.style={state, 
                                    minimum width=1.2cm,
                                    line width=0.5mm,
                                    fill=white!20!white}}
 
        \node[node style, initial left] at (0, 0)  (s0)     {\Large $rm_0$};
        \node[node style] at (6, 0)  (s1)     {\Large $rm_1$};

        \draw[every loop,
              auto=right,
              line width=0.5mm,
              >=latex,
              draw=black,
              fill=black]
            (s0)     edge[loop above, auto=left] node {\texttt{inspect},thresh =0,$1$ } (s0)
             (s0)     edge[loop below, auto=left] node {\texttt{check\_maintenance}, trig =0,$1$ } (s0)
            (s0)     edge[bend left=20, auto=left] node {\texttt{check\_maintenance}, trig=1,$1$} (s1)
            (s0)     edge[bend left=00, auto=left] node {\texttt{inspect}, thresh =1,$1$} (s1)
            (s1)     edge[bend left=20, auto=left] node {\texttt{trigger\_maintenance},$1$} (s0);
           
    \end{tikzpicture}}
		\caption{{CTMC representing the RM with $\Act_{RM} = \{ \texttt{inspect}, \texttt{check\_maintenance}, \texttt{perform\_maintenance} \}$ shown on the state transitions. The guard condition $\mathit{trig} = 0 / 1$ or $\mathit{thresh} = 0/1$ must be satisfied for the corresponding transition to trigger when it is activated via synchronisation with the transition label.}}\label{fig:RM}
	\end{subfigure}
	\\
	\begin{subfigure}{\textwidth}
		\centering
		\resizebox{0.7\textwidth}{!}{    \begin{tikzpicture}[font=\sffamily]
 
        \tikzset{node style/.style={state, 
                                    minimum width=1.2cm,
                                    line width=0.5mm,
                                    fill=white!20!white}}
 
        \node[node style, initial left] at (0, 0)  (s0)     {\Large $im_0$};
        \node[node style] at (6, 0)  (s1)     {\Large $im_1$};

        \draw[every loop,
              auto=right,
              line width=0.5mm,
              >=latex,
              draw=black,
              fill=black]
            (s0)     edge[loop above, auto=left] node {\texttt{inspect}, thresh =0,$1$} (s0)
            (s0)     edge[bend left=00, auto=left] node {\texttt{inspect}, thresh =1,$1$} (s1)
            (s1)     edge[bend left=30, auto=left] node {\texttt{perform\_maintenance} ,$1$} (s0);
           
    \end{tikzpicture}}
		\caption{{CTMC representing the IM with $\Act_{IM} = \{\texttt{inspect}, \texttt{perform\_maintenance}\}$ shown on the state transitions. The guard condition $\mathit{trig} = 0 $ and $\mathit{thresh} = 1$ must be satisfied for the corresponding transition to trigger when it is activated via synchronisation with the transition label.}}\label{fig:IM}
	\end{subfigure}
	\caption{{CTMC for (a) RM and (b) IM.}}
	\label{fig:CTMCRMIM}
\end{figure}
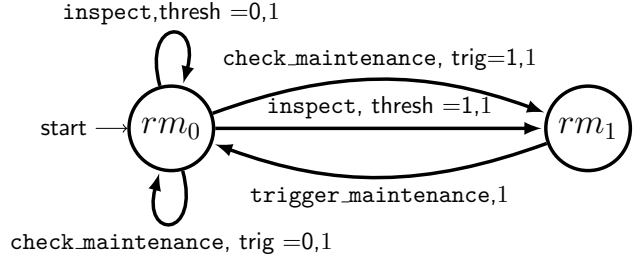
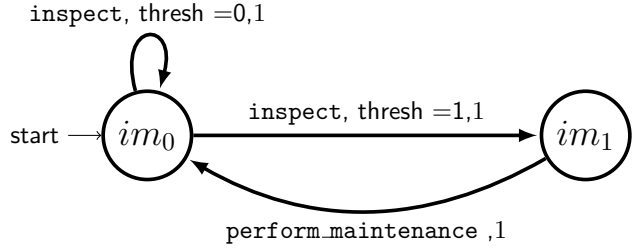

\paragraph{\textbf{Repair module (RM)}}
\label{par:FMTFrame:Modelling:RM}
Figure  \ref{fig:CTMCRMIM} \subref{fig:RM} shows the semantics of $(RM,n,$ $\Trp, \Toh,$ $ \Tin, \Tcln,\Trpl, T_{rplc},\mathit{thresh},\mathit{trig})$.    The CTMC is described using { the state space  $\{rm_0,rm_1\}$, the initial state $rm_0$, the transition labels
	$	 \Act_{RM} = \{\texttt{inspect}, $ $ \texttt{check\_clean}, \texttt{check\_replace},$ $\texttt{trigger\_clean}, $ $\texttt{trigger\_replace} \}$,
	the atomic propositions $\Ap=\{maintenance\}$, the labelling function $L(rm_{0})=\{ \emptyset \}, L(rm_1) = \{ \mathit{maintenance} \}$ and with }	$
\Rt_{IM} = \left[\begin{smallmatrix}
1&1\\
1&0\\	
\end{smallmatrix}\right]$. 
{For the sake of clarity in Figure~\ref{fig:CTMCRMIM} \subref{fig:RM}, we used the transition labels \texttt{check\_maintenance} and \texttt{trigger\_maintenance}. The transition label \texttt{check\_maintenance}  and corresponding transitions are replicated and the transition labels replaced by  \texttt{check\_clean}  or \texttt{check\_replace} to allow for both type of maintenance checks.  Similarly, the transition label \texttt{trigger\_maintenance} and corresponding transitions are duplicated and the transition labels replaced by  \texttt{trigger\_clean}  or \texttt{trigger\_replace} to allow the initiation of both type of maintenance actions to be performed. Due to synchronisation, only one of the transitions may trigger at any time instance (as explained in Subsection~\ref{subsubsec:FMTFrame:Modelling:Semantics}).	
	\noindent The transition labels   \texttt{trigger\_clean}  or \texttt{trigger\_replace} correspond to the transition label \texttt{trigger} within the DELAY module approximating the deterministic delays $\Tcln$ and $\Trpl$ respectively. 
	The deterministic delays which trigger \texttt{inspect}, \texttt{check\_clean}  or \texttt{check\_replace}  correspond to when the time delays $\Tin, \Trp$ and $\Toh$ respectively, have elapsed. All these signals are generated using individual DELAY modules with the \texttt{move} transition label for each module replaced using   \texttt{inspect}, \texttt{check\_clean}  or \texttt{check\_} \texttt{replace} respectively.
}
The $\mathit{thresh}$ signal is modelled using,
\begin{align}\label{eqn:thresh}
\mathit{thresh} &= \begin{cases}
1 &  L(s_{j,1}) = \mathit{thresh} \vee \dots \vee L(s_{j,n}) = \mathit{thresh},\\
0   & \text{otherwise},\\
\end{cases}
\end{align}
where $L(s_{j,i}), j \in 0 \dots N, i \in 1 \dots n$ correspond to  the label of the current state $j$ of each of the $n$ EBE. Similarly, we model the $\mathit{trig}$ signal using
\begin{align}\label{eqn:trig}
\mathit{trig} &= \begin{cases}
1 &  L(s_{j,1}) \neq  \mathit{new} \vee \dots \vee L(s_{j,n}) \neq \mathit{new} ,\\
0   & \text{otherwise}.\\
\end{cases}
\end{align}
Both signals act as guards which when triggered determine which transition to perform (cf. Fig.~ \ref{fig:CTMCRMIM} \subref{fig:RM}).

\paragraph{\textbf{Inspection module (IM) }}
\label{par:FMTFrame:Modelling:IM}
The semantics of the $(IM,n,\Tin,$ $ \Tcln,\Trpl,$ 
$\mathit{thresh})$ is depicted in Figure~\ref{fig:CTMCRMIM} \subref{fig:IM}.{ The CTMC is defined using the tuple $(\{im_0,im_1\},im_0,\Act_{IM}, \Ap_{IM}, L_{IM}, \Rt_{IM})$. Here,
	\begin{align*}
	\Act_{IM} = \{\texttt{inspect},\texttt{perform\_clean},\texttt{perform\_replace} \}
	\end{align*},
	 ~~ $\Ap_{IM} $ $= \{\emptyset\}$, $L(s_0)=L(s_1) = \emptyset$ and 	$
	\Rt_{IM} = \left[\begin{smallmatrix}
	1&1\\
	1&0\\	
	\end{smallmatrix}\right]$. }The $\mathit{thresh}$ signal corresponds to same signal used by the RM, given using~\eqref{eqn:thresh}. In  Figure \ref{fig:CTMCRMIM} \subref{fig:IM}, for clarity, we use the transition label \texttt{perform\_maintenance}. This transition label and corresponding transitions are duplicated and the transition labels are replaced by  either \texttt{perform\_clean}  or \texttt{perform\_replace} to allow for both type of maintenance actions to be performed when one of them is triggered using synchronisation.  The same DELAY modules used in the RM and EBE to represent the deterministic delays are used by the IM. The DELAY module used to represent the  deterministic delays $\Tcln$ and $\Trpl$ triggers the transition labels \texttt{perform\_clean}  or \texttt{perform\_replace}. This  represents that the maintenance action has completed.

%
%

\subsection{ Semantics of FMT}
\label{subsubsec:FMTFrame:Modelling:Semantics}	 
{Next, we show how to obtain the semantics of a FMT from the semantics of its elements using the FMT syntax introduced in Subsection~\ref{subsubsec:FMTFrame:Modelling:Syntax}. We define the DAG $G$ by defining the vertices $V$ and  the corresponding events $E$. The leaves of the DAG are the events corresponding to the EBE. The events $E$ are connected to the vertices $V$, which trigger the corresponding auxiliary function used to represent the semantics of the gates. The $Events$ connected to the RM and IM are initiated by triggering the auxiliary functions  $\mathit{thresh}$ and $\mathit{trig}$ given using ~\eqref{eqn:thresh} and ~\eqref{eqn:trig} respectively. Based on the structure of $G$, we compute the corresponding CTMC by applying parallel composition of the individual CTMCs representing the elements of the FMT. The parallel composition formulae are derived from~\cite{hermanns2011concurrency} and defined as follows,}
\begin{definition}[Interleaving Synchronization] \label{def:IntSynch}
	The interleaving synchronous product of $C_1 = (S_1,s_{01}, \Act_{1}, \Ap_1, L_1,\Rt_1)$ and $C_2 =(S_2,s_{02}, \Act_{2}, \Ap_2, L_2,\Rt_2)$ is $C_1 ||C_2 = (S_1 \times  S_2, (s_{01} , s_{02}),\Act_1 \cup \Act_2, \Ap_1 \cup \Ap_2, L_1 \cup L_2,\Rt)$ where $\Rt$ is given by:
	{\begin{align*}
		\frac{s_1 \xrightarrow{\alpha_1, \lambda_1} s_1'}{(s_1,s_2) \xrightarrow{\alpha_1,\lambda_1} (s_1',s_2)},~\text{and}~
		\frac{s_2 \xrightarrow{\alpha_2, \lambda_2} s_2'}{(s_1,s_2) \xrightarrow{\alpha_2,\lambda_2} (s_1,s_2')},
		\end{align*}
	}
	and $s_1,s_1' \in S_1$, $\alpha_1 \in \Act_1$, $\Rt_1(s_1,s_1') = \lambda_1$, $s_2,s_2' \in S_2$, $\alpha_2 \in \Act_2$, $\Rt_2(s_2,s_2') = \lambda_2$. 
\end{definition}
\begin{definition}[Full Synchronization]\label{def:FullSynch}
	The full synchronous product of $C_1 = (S_1,s_{01}, \Act_{1}, \Ap_1, L_1,\Rt_1)$ and $C_2 =(S_2,s_{02}, \Act_{2}, \Ap_2, L_2,\Rt_2)$ is $C_1 ||C_2 = (S_1 \times  S_2, (s_{01} , s_{02}),\Act_1 \cup \Act_2, \Ap_1 \cup \Ap_2, L_1 \cup L_2,\Rt)$ where $\Rt$ is given by:
	{\begin{align*}
		\frac{s_1 \xrightarrow{\alpha, \lambda_1} s_1' ~ \text{and}~ s_2 \xrightarrow{\alpha, \lambda_2} s_2' }{(s_1,s_2) \xrightarrow{\alpha,\lambda_1\times \lambda_2} (s_1',s_2')}
		\end{align*}
	}
	and $s_1,s_1' \in S_1$, $\alpha \in \Act_1 \wedge \Act_2$, $\Rt_1(s_1,s_1') = \lambda_1$, $s_2,s_2' \in S_2$, $\alpha_2 \in \Act_2$, $\Rt_2(s_2,s_2') = \lambda_2$. 
\end{definition}

For any pair of states, synchronisation is performed either using interleaving or full synchronisation. For full synchronisation, as in Definitions~\ref{def:IntSynch}, the rate of a synchronous transition is defined as the product of the rates for each transition. The intended rate is specified in one transition and the rate of other transition(s) is specified as 1. {For instance, the RM synchronises using full synchronisation with the DELAY modules representing $\Tin$, $\Trp$ and $\Trpl$ and therefore, to perform synchronisation between the RM and the DELAY modules, the rates of all the transitions of RM should have a value of 1 (cf. Fig.~\ref{fig:CTMCRMIM} \subref{fig:RM}), while the rate of the DELAY modules represent the actual rates (cf. Fig~\ref{fig:Clk}). The same principle holds for the EBEs and the IM. We refer the reader to Table~\ref{tab:synch} to further elucidate the synchronisation between the FMT components and the method employed during the parallel composition.} 
\begin{table*}[ht!]
	\centering
	\resizebox{\textwidth}{!}{
		\begin{tabular}{llll}
			\textbf{Component}& \textbf{Synchronised with component} & \textbf{Transition label}& \textbf{Synchronisation method}\\ \hline \hline
			DELAY representing $\Td$ &  DELAY modules representing $\Tcln,\Trpl, \Tin$ & \texttt{trigger} & Full synchronisation\\ 
			RM                       &  DELAY module representing $\Trp$ & \texttt{trigger\_clean} & Full synchronisation\\ 
			RM                       &  DELAY module representing $\Toh$ & \texttt{trigger\_replace} & Full synchronisation\\ 
			EBE & DELAY representing $\Td$ & \texttt{degrade}$_N$& Full synchronisation \\
			DELAY representing $\Tcln$  &  RM, EBE & \text{check\_clean} & Full synchronisation\\
			DELAY representing $\Trpl$  &  RM, EBE & \text{check\_replace} & Full synchronisation\\
			DELAY representing $\Tin$  &  RM, IM & \text{inspect} & Full synchronisation\\
			DELAY representing $\Trp$  &  RM, IM, EBE & \text{perform\_clean} & Full synchronisation\\
			DELAY representing $\Toh$  &  RM, IM, EBE & \text{perform\_replace} &Full synchronisation \\
			EBE                        & RM,IM, all DELAY modules, other EBEs & 	- & Interleave synchronisation\\		\hline \hline
			
	\end{tabular}}
	\caption{{Performing synchronisation between the different FMT components and the synchronisation method used.}}
	\label{tab:synch}
\end{table*}
%
\begin{example}[Synchronisation of FMT elements]
	Consider, a simple example showing the time signals and synchronisations required for modelling an 
	EBE and the RM and IM. The EBE has a degradation rate equal to $\Td$ 
	and we limit the functionality of the RM and IM by allowing only the maintenance action to perform cleaning. We also need the corresponding DELAY modules generating the degradation rates, $\Td$ and the maintenance rates $\Tcln,\Tin,\Trp$.
	The resulting CTMC is obtained by performing a parallel composition of the components $C_{all} = C_{EBE}||$ $C_{\Td}||C_{RM}||C_{IM}||C_{\Tcln}$ $||C_{\Tin}||C_{\Trp}.$ 
	\noindent The resulting state space is then $S_{all}= S_{EBE} \times S_{\Td} \times S_{RM} \times S_{IM} \times S_{\Tcln} \times S_{\Tin} \times S_{\Trp}  $. 
	The synchronisation between the different components is shown in Figure~\ref{fig:Synch} and proceeds as follows: 
	\begin{enumerate}
		\item All the DELAY modules (except $\Tcln$) start at the same time using the \texttt{trigger} transition label.
		\item When the extended DELAY module generating the $\Td$ time delay elapses, the corresponding EBE moves to the next state through synchronisation with the transition label \texttt{degrade}$_N$.
		\item The clock signals $\Trp,\Tin$ represent periodic maintenance and inspection actions and when the deterministic delay is reached, through synchronisation with the transition label \texttt{check\_clean} or the  $\texttt{inspect}$, the RM or IM modules is triggered (cf. Fig. \ref{fig:RM} and \ref{fig:IM}). If RM triggers a maintenance action, the DELAY representing $\Tcln$ is triggered using the synchronisation labels \texttt{trigger\_clean}. Once the deterministic delay $\Tcln$ elapses, the EBE, the extended DELAY module representing $\Td$ (where the \texttt{reset} transition label within the extended DELAY module is replaced with \texttt{perform\_clean} ) and the IM are reset using the transition label \texttt{perform\_clean}.
	\end{enumerate}
	
	\begin{figure}[ht!]
		\centering
		\includegraphics[width=.7\textwidth]{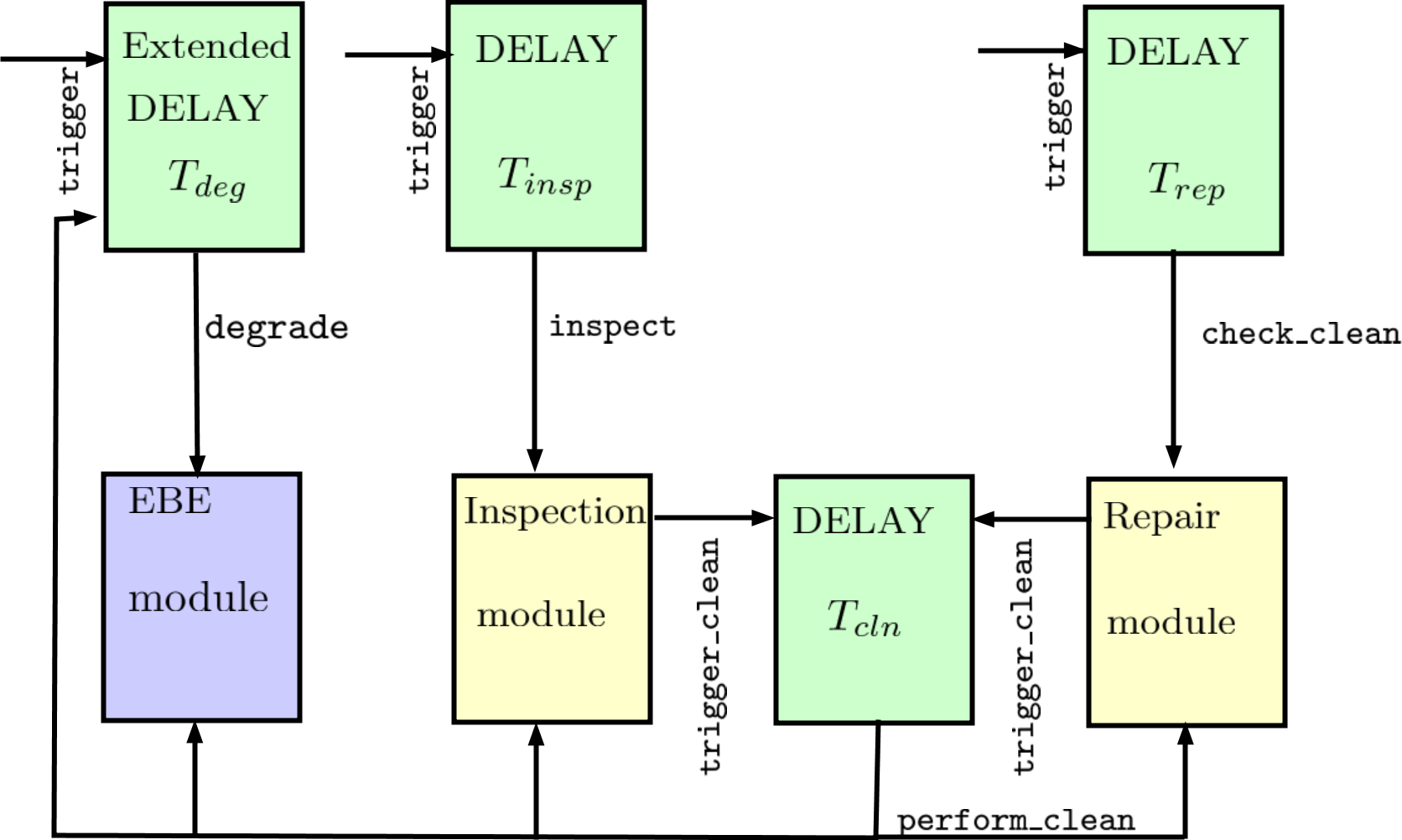}
		\caption{{Block diagram showing the synchronisation connections between one component and the other, together with the corresponding transition label which trigger synchronisation.}}
		\label{fig:Synch}
	\end{figure}
	
	\begin{remark}\label{remark:ss}
		One should note that this results in the requirement of a large state space, which is a function of the number of states used to approximate the deterministic delays. Thus, to counteract this effect we propose an abstraction framework in Subsection~\ref{subsec:FMTFrame:Dec}.
	\end{remark}

\end{example}

\subsection{Metrics}
\label{subsec:FMTFrame:Met}
We use PRISM to compute the metrics of the model described in Subsection \ref{subsec:Prelim:FMTFrame}. The metrics can be expressed using the extended Continuous Stochastic Logic (CSL) as follows:
\begin{enumerate}
	\item \textit{Reliability} :
	This can be expressed as the complement of the probability of failure over the time $T$, $1-\po_{=?}[\fo^{\le T} failed ]$.
	\item \textit{Availability}:
	This can be expressed 
	as $\ro_{=?}[C^{\le T} ]/T$, which corresponds to the cumulative reward of the total time spent in states labelled with \textit{okay} and \textit{thresh} during the time  $T$.
	\item \textit{Expected cost}:		
	This can be expressed using 
	$\ro_{=?}[C^{\le T}]$, which corresponds to the cumulative reward of the total costs (operational, maintenance and failure) within the time $T$.
	\item \textit{Expected number of failure}:		
	This can be expressed using 
	\noindent $\ro_{=?}[C^{\le T} ]$, which corresponds to the cumulative transition reward that counts the number of times the top event enters the \emph{failed state} within the time $T$.
	
\end{enumerate}

\subsection{Decomposition of FMTs}
\label{subsec:FMTFrame:Dec}
\begin{figure*}[ht!]
	\psfragscanon
	\includegraphics[width=\textwidth]{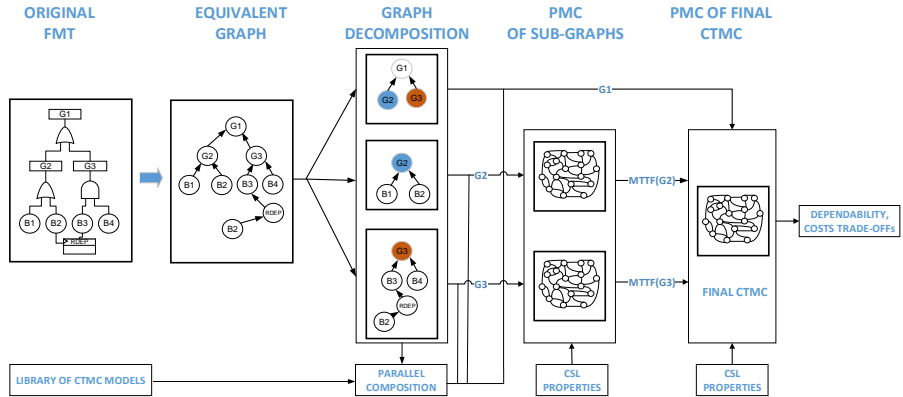}
	\caption{{Overall developed framework for decomposition of FMTs into the equivalent abstract CTMCs.}}
	\label{fig:Overall}
\end{figure*}

The use of CTMC and deterministic time delays results in the requirement of a large state space  for modelling the whole FMT (cf. Remark~\ref{remark:ss}). We therefore propose an approach which decomposes the large FMT into an equivalent abstract CTMC which can be analysed using PRISM. The process involves two transformation steps. First we convert the FMT into the equivalent directed acyclic graph (DAG) and split this graph into a set of smaller sub-graphs. Second, we transform the sub-graphs into the equivalent CTMC by making use of the developed FMT components semantics (cf. Subsec.~\ref{subsubsec:FMTFrame:Modelling:SemanticsElements}), and performing parallel composition of the individual FMT components based on the underlying structure of the sub-graph. The smaller sub-graphs are then sequentially recomposed to generate the higher level abstract FMT. Figure \ref{fig:Overall} depicts a high-level diagram of the decomposition procedure.

\paragraph{\textbf{Conversion of original FMT to the equivalent graph}}
The FMT is a DAG (cf. Subsection~\ref{subsec:FMTFrame:Modelling}) and in this framework we need to apply a transformation to the DAG in the presence of an RDEP gate, such that we can perform the decomposition. The RDEP causes an acceleration of events on dependent child nodes when the input node fails. In order to capture this feature in a DAG,  we need to duplicate the input node such that it is connected directly to the RDEP vertex. This allows us to capture when the failure of the input occurs and the corresponding acceleration of the the children.  This is reasonable as the same RM and IM are used irrespective of the underlying FMT structure.	

%
\paragraph{\textbf{Graph decomposition}}
We define modules within the DAG as sub-trees composed of at least two events which have no inputs from the rest of the tree and no outputs to the rest except from its output event~\cite{Li20151400}.	
We can divide the graph into multiple partitions based on the number of modules making up the DAG.
We define the following notations to ease in the description of the algorithm:
\begin{itemize}
	\item $V_o$ indicates whether the node is the top node of the DAG. 
	\item $V_g$ indicates the node where graph split is performed.	
	\item Modules correspond to sub-graphs in DAG.
\end{itemize} 
We set $V_o$ when we construct the DAG from the FMT and then proceed with executing Algorithm \ref{alg:split}. We first identify all the sub-graphs within the whole DAG and label all the top nodes of each sub-graph $i$ as $V_{Ti}$. We loop through each sub-graph and its immediate child (the sub-graph at immediate lower level) and at the point where the sub-graph and child are connected, the two graphs are split and a new node $V_g$ is introduced. Thus, executing Algorithm \ref{alg:split} results in a set of sub-graphs linked together by the labelled nodes $V_g$. For each of lower level sub-graphs we now proceed to compute the mean time to failure (MTTF). This will serve as an input to the higher-level sub-graphs such that metrics for the abstract equivalent CTMC can be computed.
\begin{pseudocode}
	\caption{\small{DAG decomposition algorithm}}
	\label{alg:split}
	\DontPrintSemicolon
	\LinesNumbered
	\SetKwInOut{Input}{input}\SetKwInOut{Output}{output}
	\label{alg:ADP_Nat1}
	\Input{ DAG $G=(V,E)$ }
	\Output{Set of sub-graphs with one of the end nodes labelled as $V_g$.}
	Identify sub-graphs using `depth-first' traversal \;
	Label all top nodes of each sub-graph $i$ as $V_{T_i}$ \;
	\ForAll{select the top node of every sub-graph and immediate child defined at immediate lower level }{	
		\If{label $V_T$ already found in one of the leaf nodes of sub-graph}{Split sub-graph \;
			Insert new node $V_{g}$ which will be used as input from connected sub-graph }
	}
\end{pseudocode} 

\paragraph{\textbf{PMC of sub-graphs}}

We start from the bottom level sub-graphs and perform the conversion to CTMC using the formal models presented in Subsection~\ref{subsubsec:FMTFrame:Modelling:SemanticsElements}. The formal models have been built into a library of PRISM modules and based on the underlying components and structure making up the sub-graph, the corresponding individual formal models are converted into the sub-graph's equivalent CTMC by performing parallel composition (cf.  Subsec. \ref{subsubsec:FMTFrame:Modelling:Semantics}). For each sub-graph, we compute the probability of failure $D_e(T)$ at time $T$
, from which we calculate the MTTF using,
$\mathit{MTTF} = \frac{\ln(1-D_e(T))}{-T}$.
The MTTF serves as the input to the higher level sub-graph at time $T$. The new node in the higher-level sub-graph, now degrades with the a new time delay $\Td = \mathit{MTTF}$, which is fed into the corresponding DELAY component. This process is repeated for all the different sub-graphs until the top level node $V_o$ is reached.

\begin{figure}[ht!]
	\centering
	\includegraphics[width=0.8\textwidth]{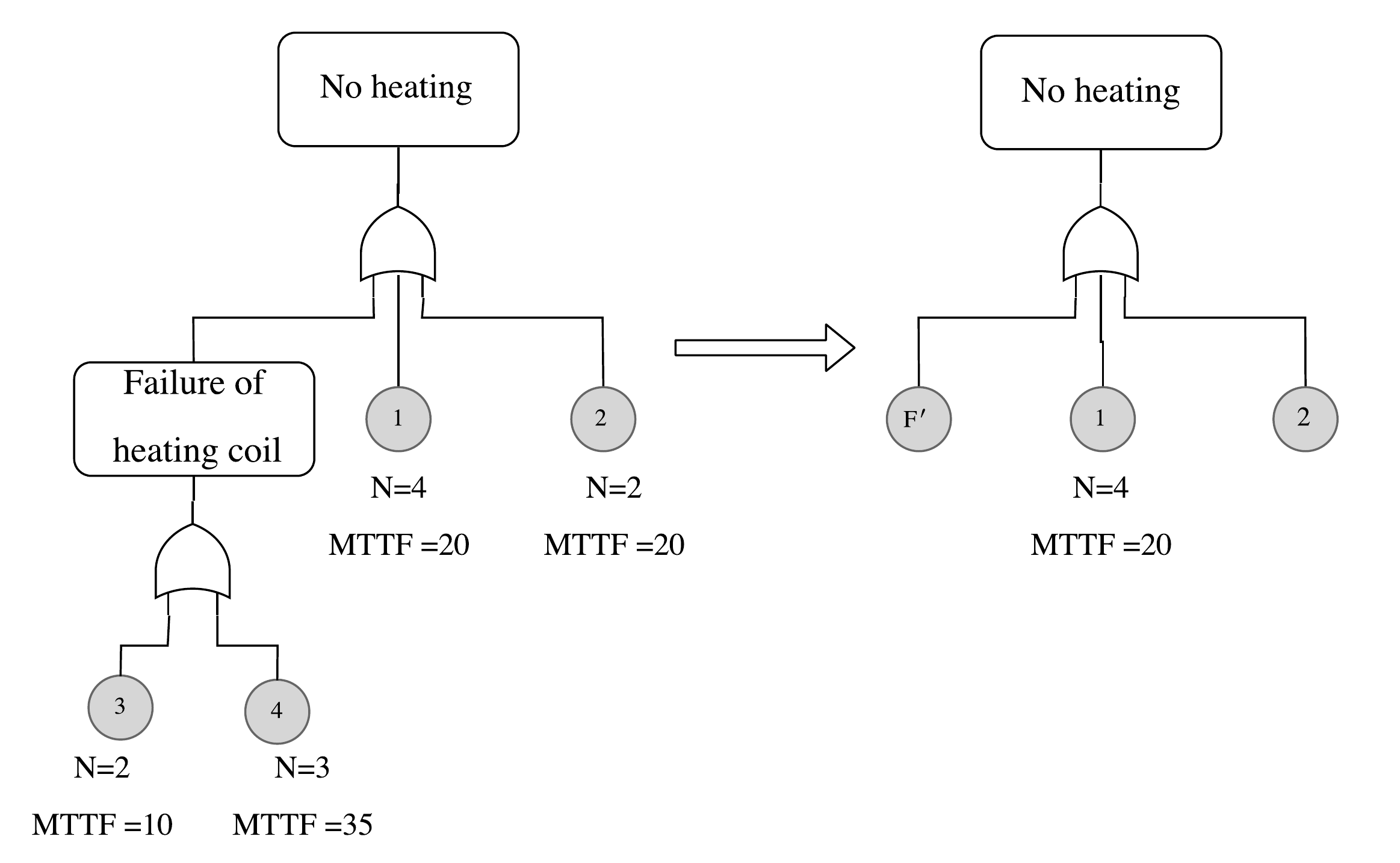}
	\caption{{The original FMT and the abstract FMT corresponding to the equivalent abstract CTMC generated by the developed framework. The MTTF for the F' is computed based on the probability of failure of the heating coil.}}
	\label{fig:IncVer}
\end{figure}

\paragraph{\textbf{PMC of final equivalent abstract CTMC}}
On reaching the top level node $V_o$, we compute the metrics for the equivalent abstract CTMC for a specific time horizon $T$. For different horizons, the previous step of computing the MTTF for the underlying lower level sub-graphs needs to be repeated. 
Using this technique, we can formally verify larger FMTs, while using less memory and computational time due to significantly smaller state space  of the underlying CTMCs.	
Next, we proceed with an illustrative example comparing the process of directly modelling the large FMT using CTMCs versus the de-compositional modelling procedure. Figure \ref{fig:IncVer} presents the FMT composed of two modules and the corresponding abstracted FMT. The abstract FMT is a pictorial representation of the moel represented by the equivalent abstract CTMC obtained using the developed decomposition framework (cf. Fig. \ref{fig:Overall}). 
For both the large FMT and the equivalent abstract FMT a comparison between the total number of states for the resulting CTMC models, the total time to compute the reliability metric and the resulting reliability metric is performed. All computations are run on an 2.3 GHz Intel Core i5 processor  with 8GB of RAM and the resulting statistics are listed in Table \ref{tab:HVAC_SizeEg}. The original FMT has a state space  with 193543 states, while the equivalent abstract CTMC has a state space  with 63937 states. This corresponds to a $67\%$ reduction in the state space  size. The total time to compute the reliability metric is a function of the final time horizon and a maximal $73\%$ reduction in computation time is achieved. Accuracy in the reliability metric of the abstract model is a function of the time horizon. 
The accuracy of the reliability metric computed by the abstract FMT results in a maximal reduction of $0.61\%$.
\begin{table}[ht!]
	\centering
	\resizebox{0.8\textwidth}{!}{
		\begin{tabular}{ccccccc} 
			\hline
			{\textbf{Time}}& \multicolumn{2}{c}{\textbf{Original FMT}}  &  \multicolumn{4}{c}{\textbf{Abstracted FMT}} \\ 
			\textbf{Horizon}
			&{Time to compute} 		  & Reliability
			& \multicolumn{2}{c}{Time to compute}& Total  & Reliability\\    		
			&{metric} &             
			& MTTF   & {metric} & Time &\\
			(years) & (mins)   &      & (mins) & (mins) & (mins) & 			\\ \hline \hline
			5 				
			&\textbf{0.727}   & 0.9842      
			& 0.142   & 0.181  & \textbf{0.223}     & 0.9842\\ 
			10				
			&\textbf{1.406}   & 0.8761     
			& 0.219   & 0.309  & \textbf{0.528 }   & 0.8769\\
			15				
			&\textbf{2.489}   & 0.3290      
			& 0.292   & 0.622 & \textbf{0.914 }   & 0.3270 \\ \hline
	\end{tabular}}
	\caption{{Comparison between the original large FMT and the abstracted FMT.}}
	\label{tab:HVAC_SizeEg}
\end{table}	
\section{Case study}
\label{sec:CaseStudy}
\begin{figure*}[ht!]
	\centering
	\includegraphics[width=\textwidth]{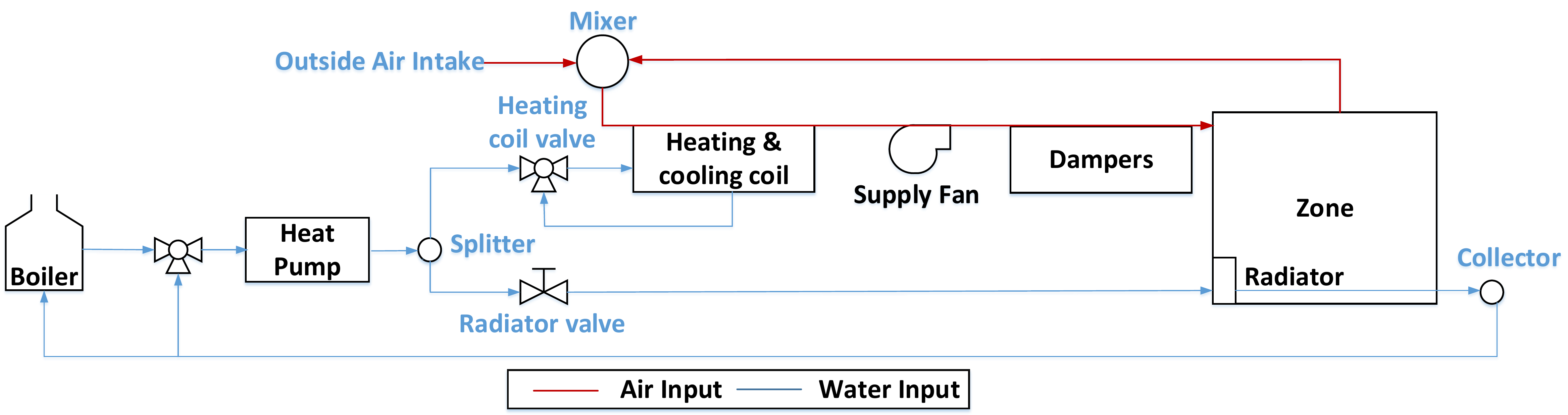}
	\caption{{High level schematic of an  HVAC system.}}
	\label{fig:hvac-system}
\end{figure*}

\begin{figure}[ht!]
	\centering
	\includegraphics[width=.8\textwidth]{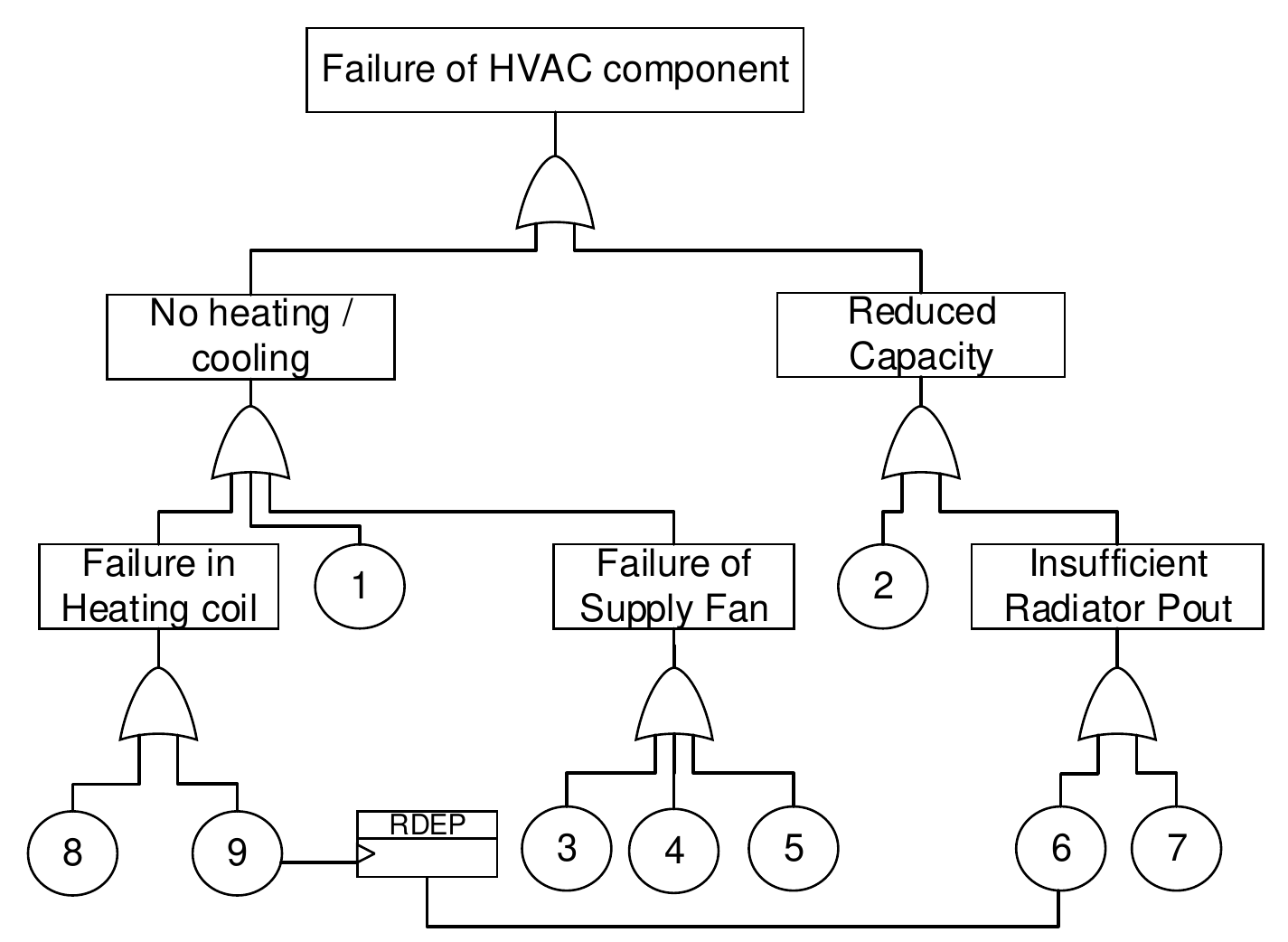}
	\caption{{FMT for failure in HVAC system with leaves represented using EBE (associated RM and IM not shown in figure). The EBE are labelled to correspond to the component failure they represent using the fault index presented in Table~\ref{tab:HVAC_FM}.}}
	\label{fig:FMT}
\end{figure}

We apply the FMT framework to a Heating, Ventilation and Air-conditioning (HVAC) system used to regulate a building's internal environment. The HVAC system under consideration for the FMT analysis is presented in Figure \ref{fig:hvac-system}.  It is composed of two circuits - the air flow circuitry and the water circuit. The gas boiler heats up the supply water which is fed into the heat pump. The heat pump transfers the supply water into two sections - the supply air heating and cooling coils and the radiators - via the splitter. The rate of water flowing in the heating coil is controlled using a heating coil valve, while the rate of water flow in the radiator is controlled using a separate valve. The outside air is mixed with the extracted room air temperature via the mixer. This is fed into the heating coil, which warms up the input air to the desired supply air temperature. This air is supplied back, at a rate controlled by the Air Handling unit (AHU) dampers, into the zone via the supply fan.  The radiators are directly connected to the water circuitry and transfer the heat from the water into the zone. The return water is then  passed through the collector and is returned back to the boiler. 
Based on this HVAC system we construct the corresponding FMT shown in Figure \ref{fig:FMT}. The leaves of the tree are EBE with discrete degradation rates computed using Table \ref{tab:HVAC_FM}, approximated by the Erlang distribution where $N$ is the number of degradation phases ($k=N$ for the Erlang distribution) and MTTF is the expected time to failure with $MTTF = 1/\lambda$ (cf. Remark~\ref{remark:Er}). We choose an acceleration factor $\gamma = 2$ for the RDEP gate. The system is periodically repaired every 6 months ($\Trp = 182\;days$) and a major overhaul with a complete replacement of all components is carried out once every 20 years ($\Toh = 20\times 365\;days$). Weekly inspections are performed ($\Tin = 7\;days$) which return the components back to the previous state. Only cleaning actions are performed when inspections are carried out. The total time to perform a cleaning action is 1 day ($\Tcln= 1\;day$), while performing a total replacement of components takes 7 days ($\Trpl = 7\;days$). The time timing signals $\{\Trp, \Toh,\Tin,\Tcln,\Trpl\}$ are all approximated using the Erlang distribution with $N = 3$. All maintenance actions are performed simultaneously on all components. 
\begin{table}[ht!]
	\centering
		\begin{tabular}{*5c} 
			\hline
			\textbf{Fault Index } & \textbf{Failure Mode} & \textbf{N} & \textbf{MTTF}\\ 
			& 					    &  &\textbf{(years)} \\ \hline \hline
			1 & Failure in cooling coil & 4&20  \\ 
			2 & Broken AHU Damper & 2 & 20 \\ 
			3 & Fan motor failure & 3 & 35 \\ 
			4 & Obstructed supply fan& 4 & 31 \\
			5 & Fan bearing failure &6 &17 \\ 
			6 & Radiator failure & 4 & 25 \\ 
			7 & Radiator stuck valve & 2 & 10 \\  
			8 & Heater stuck valve & 2&10 \\ 
			9 & Failure in heat pump &4&20 \\ \hline
	\end{tabular}
	\caption{{Extended Basic events in FMT with associated degradation rates (N, MTTF) obtained from  ~\cite{handbook1996hvac,Khan2003561}.}}
	\label{tab:HVAC_FM}
\end{table}  
\subsection{Quantitative results}
\label{subsec:CaseStudy:Res}
\begin{figure}[ht!]
	\psfragscanon
	\begin{subfigure}{0.5\textwidth}		
		\resizebox{.9\textwidth}{!}{
%
%
\definecolor{mycolor1}{rgb}{0.00000,0.44700,0.74100}%
\definecolor{mycolor2}{rgb}{0.85000,0.32500,0.09800}%
\definecolor{mycolor3}{rgb}{0.92900,0.69400,0.12500}%
\definecolor{mycolor4}{rgb}{0.49400,0.18400,0.55600}%
\definecolor{mycolor5}{rgb}{0.46600,0.67400,0.18800}%
\definecolor{mycolor6}{rgb}{0.30100,0.74500,0.93300}%
\definecolor{mycolor7}{rgb}{0.63500,0.07800,0.18400}%
\begin{tikzpicture}

\begin{axis}[%
width=1.5in,
height=.8in,
at={(0in,0in)},
scale only axis,
xmin=0,
xmax=25,
xmajorgrids,
xlabel={\footnotesize Time (Years)},
ymin=0,
ymax=1,
ymajorgrids,
ylabel={\footnotesize 1-$\po_{=?}[\diamond^{\le T} failed ]$},
axis background/.style={fill=white},
title style={font=\bfseries},
legend style={legend cell align=left,align=left,draw=white!15!black}
]
\addplot [color=mycolor1,solid,line width=1.5pt]
  table[row sep=crcr]{%
0	1\\
5	0.9963\\
10	0.7796\\
15	0.4764\\
20	0.2558\\
25	0.1167\\
};
\end{axis}
\end{tikzpicture}%
}
		\caption{{Reliability of HVAC system.}}\label{fig:Rel}
	\end{subfigure}
	\begin{subfigure}{0.5\textwidth}
		\resizebox{.9\textwidth}{!}{	
%
%
\definecolor{mycolor1}{rgb}{0.00000,0.44700,0.74100}%
\definecolor{mycolor2}{rgb}{0.85000,0.32500,0.09800}%
\definecolor{mycolor3}{rgb}{0.92900,0.69400,0.12500}%
\definecolor{mycolor4}{rgb}{0.49400,0.18400,0.55600}%
\definecolor{mycolor5}{rgb}{0.46600,0.67400,0.18800}%
\definecolor{mycolor6}{rgb}{0.30100,0.74500,0.93300}%
\definecolor{mycolor7}{rgb}{0.63500,0.07800,0.18400}%
\begin{tikzpicture}

\begin{axis}[%
width=1.5in,
height=.8in,
at={(0in,0in)},
scale only axis,
xmin=0,
xmax=25,
xmajorgrids,
xlabel={\footnotesize Time (Years)},
ymin=0.4,
ymax=1,
ymajorgrids,
ylabel={\footnotesize $\ro_{\{``Avail''\}=?}[C^{\le T} ]/T$},
axis background/.style={fill=white},
title style={font=\bfseries},
legend style={legend cell align=left,align=left,draw=white!15!black}
]
\addplot [color=mycolor1,solid,line width=1.5pt]
  table[row sep=crcr]{%
0	1\\
5	0.940298458\\
10	0.885243412\\
15	0.70616614\\
20	0.547515414\\
25	0.411231232\\
};
\end{axis}
\end{tikzpicture}%
}
		\caption{{Availability of HVAC system.}}\label{fig:Avail}
	\end{subfigure}
	\caption{{Reliability and availability of HVAC over time horizon $N_r$.}}
	\label{fig:Rel_Avail}
\end{figure}
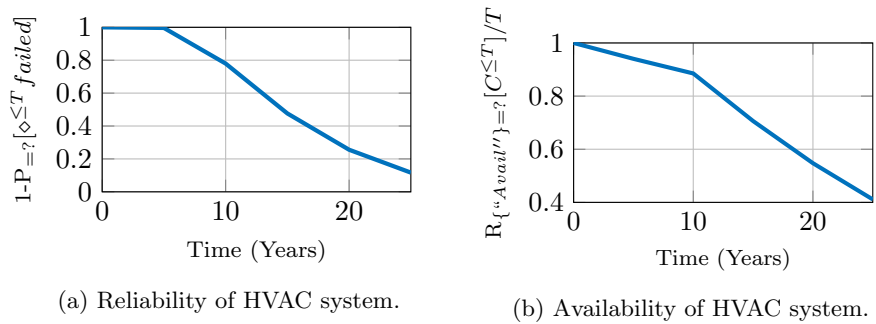

We make use of the developed framework (cf. Subsec. \ref{subsec:FMTFrame:Dec}) and convert the FMT representing the failure of the HVAC system (cf. Fig. \ref{fig:FMT}) into the equivalent abstract CTMC. The abstracted CTMC has a state space  of 62779 states. Using our current computing set-up, the complex CTMC representing the whole FMT was not computable as it results in a state space  explosion. Highlighting, the advantage of the developed framework. The process is performed over six time horizons $N_r= \{0,5,10,15,20,$ $25\}$ years with the maintenance policy consisting of periodic cleaning every 6 months, a major overhaul 
every 20 years and inspections 
on a weekly basis. For this set-up, the metrics corresponding to the reliability and availability of the HVAC systems over the time horizon are computed and are shown in Figure \ref{fig:Avail}. The maximal time taken to compute a metric using the abstract FMT is 1.47 minutes. It is deduced that both the reliability and availability reduce over time and there is a saturation in the number of maintenance actions which one can perform before the system no longer achieves higher performance in reliability and availability.
Next, we compare the total cost of maintenance and the expected number of failures over the time horizon $N_r= \{0,5,10,15,20,25\}$ years when considering different maintenance strategies, such that we can identify the maintenance strategy that minimises cost and the number of failures over time. We consider six different maintenance strategies which are listed in Table \ref{tab:MaintStrat}. The total maintenance cost to perform a repair is 100 [GBP], while a replacement costs 5000 [GBP]. We now compute the total expected maintenance costs and the total expected number of failures for each strategy. These are shown in Figure \ref{fig:NumFailures}. The most effective strategy which offers a good trade-off between maintenance costs and the expected number of failures is achieved when repairs are carried out on a yearly basis, replacements are carried out every 20 years and inspections are carried out weekly (corresponding to strategy $M_1$). Furthermore, it can be seen that the frequency of inspections has a large effect on the total number of failures. When the frequency of inspection is low (as in $M_4$ and $M_5$), the expected number of component failures increases significantly. Note that reducing the periodicity of repairs, as in the case of maintenance strategy $M_2$ also results in an increase in the expected number of failures.
\begin{table}[ht!]
	\centering
	\begin{tabular}{*5c} 
		\hline
		\textbf{Strategy index} & {$\Trp$} & $\Toh$ & $\Tin$\\ \hline \hline
		$M_0$ & 6 months   & 20 years & 1 Week  \\ 
		$M_1$ & 12 months  & 20 years & 1 Week  \\ 
		$M_2$ & 48 months  & 20 years & 1 Week  \\ 
		$M_3$ & 6 months   & 10 years & 1 Week \\
		$M_4$ & 6 months   & 20 years & 2 years \\ 
		$M_5$ & 6 months   & 20 years & 5 years \\   \hline
	\end{tabular}
	\caption{{Implemented maintenance strategies}}
	\label{tab:MaintStrat}
\end{table}	
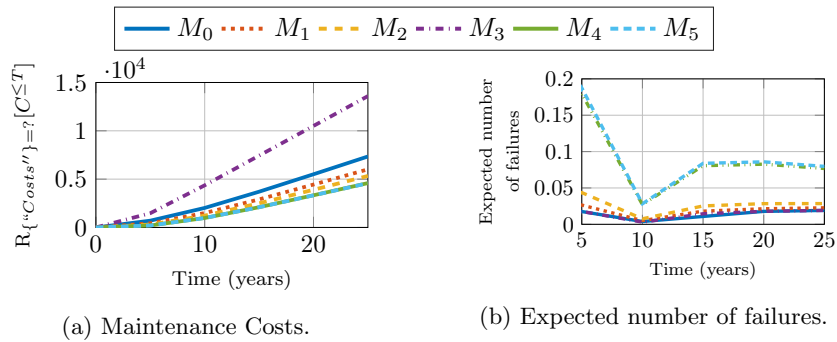
\begin{figure}[h!]
	\psfragscanon \centering
	\begin{tikzpicture}
	\begin{axis}[%
	hide axis,
	xmin=10,
	xmax=50,
	ymin=0,
	ymax=0.4,
	legend style={at={(0,0)},anchor=north,legend columns=6,legend cell align=left,align=left,draw=white!15!black,legend cell align=left}
	]
	\addlegendimage{color=mycolor1,solid,line width=1.5pt}
	\addlegendentry{ $M_0$};
	\addlegendimage{color=mycolor2,dotted,line width=1.5pt}
	\addlegendentry{ $M_1$};
	\addlegendimage{color=mycolor3,dashed,line width=1.5pt}
	\addlegendentry{ $M_2$};
	\addlegendimage{color=mycolor4,dash dot,line width=1.5pt}
	\addlegendentry{ $M_3$};
	\addlegendimage{color=mycolor5,solid,line width=1.5pt}
	\addlegendentry{ $M_4$};
	\addlegendimage{color=mycolor6,densely dashed,line width=1.5pt}
	\addlegendentry{ $M_5$};
	\end{axis}
	\end{tikzpicture}%
	\\
	\begin{subfigure}{0.5\textwidth}
		\centering
		\resizebox{.8\textwidth}{!}{
%
%
\definecolor{mycolor1}{rgb}{0.00000,0.44700,0.74100}%
\definecolor{mycolor2}{rgb}{0.85000,0.32500,0.09800}%
\definecolor{mycolor3}{rgb}{0.92900,0.69400,0.12500}%
\definecolor{mycolor4}{rgb}{0.49400,0.18400,0.55600}%
\definecolor{mycolor5}{rgb}{0.46600,0.67400,0.18800}%
\definecolor{mycolor6}{rgb}{0.30100,0.74500,0.93300}%
\definecolor{mycolor7}{rgb}{0.63500,0.07800,0.18400}%
\begin{tikzpicture}
\begin{axis}[%
width=1.5in,
height=.8in,
at={(0in,0in)},
scale only axis,
xmin=0,
xmax=25,
xmajorgrids,
xlabel={\footnotesize Time (years)},
ymin=0,
ymax=15000,
ymajorgrids,
ylabel={\footnotesize $\ro_{\{``Costs''\}=?}[C^{\le T} ]$},
axis background/.style={fill=white},
title style={font=\bfseries},
]
\addplot [color=mycolor1,solid,line width=1.5pt]
  table[row sep=crcr]{%
0   0\\
5	701.6556158\\
10	2020.713667\\
15	3700.034496\\
20	5503.600467\\
25	7343.398342\\
};
\addplot [color=mycolor2,dotted,line width=1.5pt]
table[row sep=crcr]{%
0   0\\
5	437.1184346\\
10	1490.970659\\
15	2903.883752\\
20	4440.156867\\
25	6011.22231\\
};
\addplot [color=mycolor3,dashed,line width=1.5pt]
table[row sep=crcr]{%
0   0\\
5	304.4261632\\
10	1224.161839\\
15	2500.625101\\
20	3898.959974\\
25	5329.707406\\
};
\addplot [color=mycolor4,dash dot,line width=1.5pt]
table[row sep=crcr]{%
0   0\\
5	1473.40538\\
10	4361.247102\\
15	7432.879357\\
20	10506.96387\\
25	13591.52356\\
};
\addplot [color=mycolor5,solid,line width=1.5pt]
table[row sep=crcr]{%
0   0\\
5	202.9257369\\
10	978.9741113\\
15	2096.467399\\
20	3330.594595\\
25	4589.469963\\
};
\addplot [color=mycolor6,densely dashed,line width=1.5pt]
table[row sep=crcr]{%
0	0\\
5	202.3762797\\
10	976.8227041\\
15	2091.7584\\
20	3322.590937\\
25	4577.594535\\
};
\end{axis}
\end{tikzpicture}%
}
		\caption{{Maintenance Costs.}}\label{fig:Costs}
	\end{subfigure}~
	\begin{subfigure}{0.5\textwidth}
		\centering
		\resizebox{.83\textwidth}{!}{	
%
%
\definecolor{mycolor1}{rgb}{0.00000,0.44700,0.74100}%
\definecolor{mycolor2}{rgb}{0.85000,0.32500,0.09800}%
\definecolor{mycolor3}{rgb}{0.92900,0.69400,0.12500}%
\definecolor{mycolor4}{rgb}{0.49400,0.18400,0.55600}%
\definecolor{mycolor5}{rgb}{0.46600,0.67400,0.18800}%
\definecolor{mycolor6}{rgb}{0.30100,0.74500,0.93300}%
\definecolor{mycolor7}{rgb}{0.63500,0.07800,0.18400}%

\begin{tikzpicture}

\begin{axis}[%
width=1.5in,
height=.9in,
at={(0in,0in)},
scale only axis,
xmin=5,
xmax=25,
xmajorgrids,
xlabel={\footnotesize Time (years)},
ymin=0,
ymax=0.2,
ymajorgrids,
ylabel style={align=center}, ylabel=\footnotesize Expected number \\ \footnotesize of failures,
axis background/.style={fill=white},
title style={font=\bfseries},
yticklabels={0,0, 0.05, 0.1,0.15,0.2}]
\addplot [color=mycolor1,solid,line width=1.5pt]
table[row sep=crcr]{%
	0 0\\
	5 0.0176\\
	10 0.00383611
	15 0.0145\\
	20 0.0178\\
	25 0.0191\\
};
\addplot [color=mycolor2,dotted,line width=1.5pt]
table[row sep=crcr]{%
	0 0\\
	5 0.0265\\
	10 0.0051\\
	15 0.0181\\
	20 0.0214\\
	25 0.02268\\
};
\addplot [color=mycolor3,dashed,line width=1.5pt]
table[row sep=crcr]{%
	0 0\\
	5 0.04368\\
	10 0.0076\\
	15 0.025288\\
	20 0.02842\\
	25 0.028521\\
};
\addplot [color=mycolor4,dash dot,line width=1.5pt]
table[row sep=crcr]{%
	0 0\\
	5 0.01769\\
	10 0.003\\
	15 0.0145\\
	20 0.01787\\
	25 0.01912\\
};
\addplot [color=mycolor5,dash dot dot,line width=1.5pt]
table[row sep=crcr]{%
	0 0\\
	5 0.17961\\
	10 0.0271\\
	15 0.0806\\
	20 0.0827\\
	25 0.0768\\
};
\addplot [color=mycolor6,densely dashed,line width=1.5pt]
table[row sep=crcr]{%
	0 0\\
	5 0.1876\\
	10 0.0282\\
	15 0.08386\\
	20 0.0859\\
	25 0.079675\\
};
\end{axis}
\end{tikzpicture}%
}
		\caption{{Expected number of failures.}}\label{fig:ETTF}
	\end{subfigure}
	\caption{{Comparison between different number of maintenance strategies for an HVAC systems. }}
	\label{fig:NumFailures}
\end{figure}
\section{Conclusion and Future Works}
\label{sec:Conc}
The paper has presented a methodology for applying probabilistic model checking to FMTs. The FMTs are modelled in the form of CTMCs which simplifies the transformation of FMT into formal models that can be analysed using PRISM. A novel technique for abstracting the equivalent CTMC model is also presented. The novel decomposition procedure tackles the issue of state space  explosion and results in a significant reduction in both the state space  size and the total time required to compute metrics.  The framework has been applied to an HVAC system and the effect of applying different maintenance strategies has been presented.  The presented framework can be further enhanced by adding more gates to the PRISM modules library which include the Priority-AND, INHIBIT, k/N gates and to incorporate lumping of states  as in ~\cite{yevkin2015efficient}, such that the state space  can be further reduced. 
\bibliographystyle{plain}
\bibliography{FMTArxiv2}

\end{document}